# Evaluating Feasibility within Power Flow

Marko Jereminov[1,2], *Graduate Student Member, IEEE*, David M. Bromberg[2], *Member, IEEE*, Amritanshu Pandey[1], *Graduate Student Member, IEEE*, Martin R. Wagner[1], *Graduate Student Member*, *IEEE*, and Larry Pileggi[1], *Fellow, IEEE*

*Abstract*—Recent development of techniques that improve the convergence properties of power flow simulation have been demonstrated to facilitate scaling to large system sizes (80k+ buses). However, the problem remains to reliably identify cases that are infeasible – system configurations that have no solution. In this paper, we use the circuit theoretic approach based on adjoint networks to evaluate the feasibility of a power flow test case and further locate and quantify the source of infeasibility in the cases operating beyond the tip of the nose curve. By creating infeasibility current source models that are added to each node of the system model and further coupling each source to its corresponding node of the adjoint network, any locations of insufficient real or reactive power are captured by a non-zero response of the adjoint network. Furthermore, it is shown that the proposed joint simulation of power flow and its adjoint network models provide the optimally minimized currents that can be later utilized to inform corrective actions to restore the feasibility of power flow problems.

## I. INTRODUCTION

Alternating Current Power Flow (AC-PF) is a nonlinear problem that determines the steady-state operating point of the power system at a fixed frequency, and as such it represents a fundamental component in everyday operation and planning of electrical power systems. Despite the lack of convergence robustness [1], 'PQV' based AC-PF remains the industry standard for the transmission level power grid steady-state analyses. In contrast to the actual power system, where the grid frequency changes slightly with a demand change, and control systems adjust the generated power to maintain a frequency close to the nominal one [2], the corresponding power flow problem generally incorporates one or more slack bus generators to provide the additional power that is needed. Even when the real and reactive powers supplied by the slack bus are unbounded, it is still possible that no feasible solution exists due to the nonlinearities in the formulation and network topology constraints that are observed when the power network requires operation beyond the tip of the nose curve, i.e. the point of power flow Jacobian matrix singularity.

In general network theory, detailed understanding of the true physical characteristics of a network model has facilitated powerful mathematical proofs and solution methods for decades. For instance, Tellegen's theorem for complex network systems is based on conservation of energy and is derived and proven from Kirchhoff's Current and Voltage Laws [3], while the circuit simulator SPICE [3] and its many derivatives can presently solve circuit problems that contain billions of nodes. In [4]-[8] we have demonstrated that the circuit-theoretic framework utilized within state-of-the-art circuit simulators can be applied to power grid analyses, specifically power flow and three-phase power flow problems. It was shown that the equivalent circuit representation of a power flow problem provides a new perspective and intuition for understanding domain specific knowledge that can be utilized to adapt and apply methods and algorithms from the circuit simulation field [9]-[10] to improve convergence properties of large power networks [6]-[8]. Furthermore, the recently introduced Equivalent Circuit Programming (ECP) framework [11] shows that embedding of domain-specific circuit simulation techniques within the line-search optimization algorithms additionally ensures the optimal Newton Raphson (NR) residual decrement, and thus improves simulation efficiency. However, since divergence cannot be avoided when the power system problem is infeasible, as shown in Fig. 1, it is difficult to distinguish systems that have diverged due to "lack of simulation robustness" from those that are "truly infeasible."

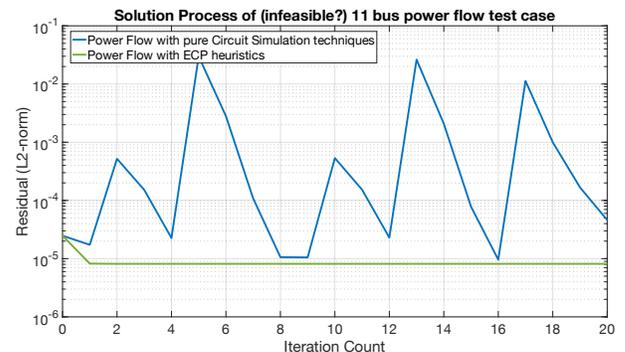

Fig. 1. Convergence profile (NR residual vs iteration count) of an infeasible 11 bus power flow test case operating at 100% loading factor. This paper will demonstrate that the case has a feasible solution at 99.82% loading factor as well as indicate the amount of power deficiency at 100% loading factor that caused the infeasibility of the test case.

There have been attempts to develop algorithms to detect power flow infeasibility. In [12], the authors discussed the conditions that define the upper bound on the number of feasible power flow solutions based on the network topology, while [13] introduces a predictor-corrector technique to explore the feasible solution space of power flow. Various homotopy methods such as the Continuation Power Flow method (CPF)

This work was supported in part by the Defense Advanced Research Projects Agency (DARPA) under award no. FA8750-17-1-0059 for the RADICS program, and the National Science Foundation (NSF) under contract no. ECCS-1800812.

[1]Authors are with the Department of Electrical and Computer Engineering, Carnegie Mellon University, Pittsburgh, PA 15213 USA (e-mail: {mjeremin, amritanp, mwagner, pileggi}@andrew.cmu.edu).

[2]Authors are with Pearl Street Technologies, (e-mail: {mjereminov, bromberg}@pearlstreettechnologies.com).

[14] have been proposed to solve the sequence of power flow problems. In [15] the two sufficient conditions for which the power flow problem does not have a solution are defined based on semidefinite relaxation of power flow and reactive power limits and further used as a feasibility metric.

One approach that attempted to identify and correct the power flow infeasibility was presented in [16], where the author formulated the power flow problem in terms of a least squares minimization to quantify the grid infeasibility. The approach is described as finding a solvable boundary and the best direction to shed the loads for restoring the feasibility, but was shown to suffer from divergence [17] and lead to non-physical local solutions [18].

Another approach to detect the infeasible power flow case is to use a provably convergent algorithm such the one introduced in [19]. As an alternative method for iteratively solving the power flow problem, the noniterative "Holomorphic Embedding Load-Flow Method" (HELM) is purported to find a correct power flow solution if one exists. However, the approach as presently described handles only PQ buses, and has issues with accurately modeling PV buses [20]. Furthermore, the presented algorithms do not seem to scale well to large power flow systems [20]. Therefore, the goal remains to create a generalized and scalable framework that not only detects infeasibility but localizes nodes/elements in the system that cause it.

In this paper we demonstrate how the adjoint network of a power system can be used for locating and evaluating power flow infeasibility. To that end, a significant contribution of this paper is the development of adjoint networks for the power flow model based on adjoint circuit theory [21]-[22] that are presented in Section III. It is shown that the power flow violations that arise for an infeasible system as a consequence of over-constraining the governing network equations can be optimally captured by the adjoint power flow network. Specifically, this is done by coupling every bus within the system to its respective bus in the adjoint network. Therefore, power flow violations can be identified by jointly solving the original power flow problem coupled with its adjoint network. Most importantly, the formulation described herein eliminates the uncertainty that arises today when a power flow simulation does not converge, considering that the solution to the jointly solved power flow and its adjoint network model *always exists*, and divergence would therefore indicate a "lack of simulation robustness." As such, it also provides an additional proof of robustness for the recently introduced circuit simulation techniques [4]-[8], considering that the source of simulation divergence can be exactly localized and quantified.

Interestingly, as shown in Section IV., *the governing equations defining the system and its adjoint network in terms of current and voltage state variables exactly represent the necessary optimality conditions of the optimization problem that minimizes the L2-norm of the additional current sources connected to each bus of the power flow network representation.* Moreover, the simulation framework comprised of jointly solving a power flow and adjoint network was recently demonstrated to represent a generic framework that can include any power flow optimization objective as shown in [11],[23]-[24].

Lastly, it is worth noting that the introduced circuit theoretic formulation for evaluating feasibility within a power flow simulation can also be solved with any of the generic state-of-the-art nonlinear equation/optimization toolboxes. However, since the generalized nonlinear toolboxes are known not to scale well with problem size, in this paper we apply domain-specific information obtained from the adjoint circuit perspective to the optimization problem to robustly solve it by extending the recently introduced circuit simulation-based power flow methods described in [6]-[8]. *It is shown that utilizing specific variable constraints and applying a unique form of homotopy, both of which are inspired by and derived from circuit models, enables a robust evaluation of feasibility for large system sizes, as well as provides simulation runtime improvements of up to 200x.* Several results are presented in Section VII to validate the proposed approach and demonstrate its practical utility for identifying and locating infeasibility.

## II. POWER FLOW CURRENT/VOLTAGE FORMULATION

Modeling the power flow problem in terms of current and voltage (I-V) state variables that can be represented by an equivalent split-circuit for the Newton-Raphson (NR) linearized equations was demonstrated to provide a generalized framework for robust and efficient power grid steady-state analyses [4]-[7]. In this section, we briefly review the split-circuit formulation concept by deriving equivalent circuit models for some of the prominent power flow models. More details and derivations of other power-flow split-circuit models can be found in [4]-[5], [25]-[26]. *It is important to emphasize that a complete linearized split-circuit power flow model represents nothing more than the physical representation of the linearized set of NR equations that is iteratively solved until convergence, as in any other nonlinear equation solver. The key difference, however, is the domain specific knowledge that can now be obtained and used for NR-step control, and as such ensures stable and efficient convergence properties.*

### A. Transmission Line π-model

Consider a series element of a transmission line $\pi$ model connecting buses $k$ and $m$. Its complex governing equation can be obtained from Ohm's Law in terms of the series line admittance $(G_L + jB_L)$ and the voltage across it $(\tilde{V}_{km})$ as:

$$\tilde{I}_{km} = (G_L + jB_L)\tilde{V}_{km} \quad (1)$$

We further split the complex current from (1) into its real and imaginary components ($I_{km}^R$ and $I_{km}^I$) as:

$$I_{km}^R = G_L V_{km}^R - B_L V_{km}^I \quad (2)$$
$$I_{km}^I = G_L V_{km}^I + B_L V_{km}^R \quad (3)$$

To represent the line π model as an equivalent split-circuit, the terms from (2)-(3), where the real and imaginary currents are proportional to the respective real and imaginary voltages across them ($V_{km}^R$ and $V_{km}^I$), define a conductance ($G_L$), while the current terms proportional to the voltage across the other circuit represent voltage-controlled current sources. After applying the same approach to map the shunt parts of π model into its split-circuit equivalent [4], the complete transmission line power flow split-circuit can be obtained as shown in Fig. 2.

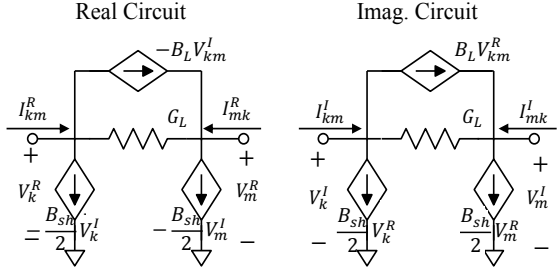

Fig. 2. Power flow split-circuit of a transmission line π-model.

### B. Modeling nonlinear constant power models

The constant power-based PV and PQ circuit models are obtained from the definition of complex power in terms of the current injection that absorbs the real ($P_C$) and reactive ($Q_C$) powers at a bus voltage ($\tilde{V}_C = V_{RC} + jV_{IC}$) given as:

$$\tilde{I}_C = \frac{P_C - jQ_C}{\tilde{V}_C^*} \quad (4)$$

where a subscript $C = \{G, L\}$ represents a placeholder denoting the current injection model corresponding to generation ($G$) or load ($L$).

To allow the application of NR, we split the nonanalytic complex current ($\tilde{I}_C$) from (4) to its real and imaginary components ($I_{RC}$ and $I_{IC}$):

$$I_{RC} = \frac{P_C V_{RC} + Q_C V_{IC}}{V_{RC}^2 + V_{IC}^2} \quad (5)$$

$$I_{IC} = \frac{P_C V_{IC} - Q_C V_{RC}}{V_{RC}^2 + V_{IC}^2} \quad (6)$$

Since the PQ load constrains both real and reactive powers, its equivalent split-circuit model [4] is defined in terms of real and imaginary load currents ($I_{RL}$ and $I_{IL}$) linearized by the first order Taylor expansion for $(k+1)^{th}$ iteration:

$$I_{RL}^{k+1} = \alpha_R^k + \frac{\partial I_{RL}^k}{\partial V_{RL}} V_{RL}^{k+1} + \frac{\partial I_{RL}^k}{\partial V_{IL}} V_{IL}^{k+1} \quad (7)$$

$$I_{IL}^{k+1} = \alpha_I^k + \frac{\partial I_{IL}^k}{\partial V_{RL}} V_{RL}^{k+1} + \frac{\partial I_{IL}^k}{\partial V_{IL}} V_{IL}^{k+1} \quad (8)$$

The current sensitivities from (7) and (8) that relate real and imaginary currents to the voltage across them ($V_{RL}^{k+1}, V_{IL}^{k+1}$) represent a conductance, while the sensitivities that are proportional to the voltage of other circuit are mapped to voltage controlled current sources. Lastly, the constant legacy terms *known from the previous NR iteration* ($\alpha_R^k$ and $\alpha_I^k$) are mapped to independent current sources, as shown in Fig. 3.

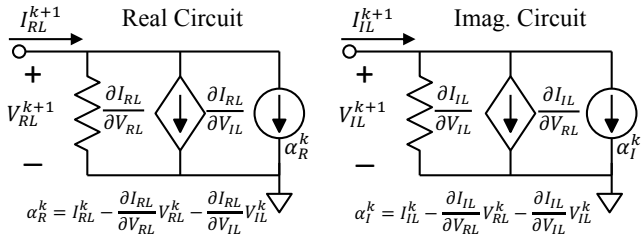

Fig. 3. Linearized power flow split-circuit of a PQ load.

In contrast to the slack bus generator whose powers are not bounded, the PV generator has a pre-set real power ($P_G$), and further adjusts its reactive power ($Q_G$) to control the bus voltage magnitude ($|V_s|$) [4]. It should be noted that the reactive power $Q_G$ represents an added state variable, hence an additional constraint that relates real and imaginary voltages across the generator ($V_{RG}, V_{IG}$) has to be added:

$$F_G \equiv V_{RG}^2 + V_{IG}^2 - |V_s|^2 = 0 \quad (9)$$

Therefore, in addition to the voltage sensitivities, the current sensitivities with respect to reactive power are added in the linearized split-circuit model [4]. Furthermore, the nonlinearities from (9) are linearized by the first order Taylor expansion and stamped (values are added to the Jacobian in a modular way) to the system of circuit equations for additional $Q_G$ variable within the formulation.

### C. Modeling Voltage Regulation (VR)

An important industry requirement represents consideration of reactive power operational limits of VR devices. Namely, a generator or other VR device can control the voltage to a prespecified set point only until one of the reactive power limits is approached. This is often referred to as PV/PQ switching, and can be defined mathematically as a disjunctive function $F_G^R$:

$$F_G^R \equiv \begin{cases} Q_{MIN} < Q_G < Q_{MAX} \wedge V_{RG}^2 + V_{IG}^2 = |V_s|^2 \\ Q_G = Q_{MAX} \wedge V_{RG}^2 + V_{IG}^2 < |V_s|^2 \\ Q_G = Q_{MIN} \wedge V_{RG}^2 + V_{IG}^2 > |V_s|^2 \end{cases} \quad (10)$$

that represents a discontinuous function characterizing the response of the VR device shown in Fig. 4.

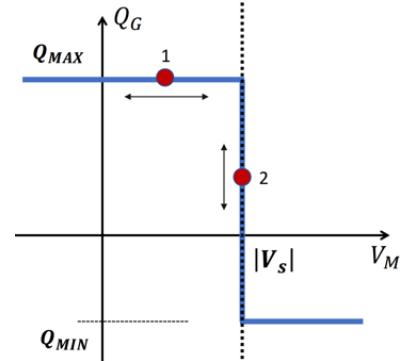

Fig. 4. Disjunctive characteristics of a voltage regulating device. Note that if generated reactive power $Q_G$ is not approaching operational limits (point 2) the voltage can be maintained at a set point, while whenever one of the limits is reached (point 1), the voltage control cannot be maintained.

The equivalent circuit formulation can include two distinct approaches for handling the disjunctive nonlinearities introduced by the characteristics of VR devices:
- *Outer control-loop (PV/PQ conversion)*

Outer control-loop approach [27] represents one of the most commonly used approaches to handle the nondifferentiable disjunctive behavior of VR devices implemented in most commercial power flow tools, and as such can be used in circuit-theoretic power flow simulators [28]. Its implementation is briefly described by the two following steps:
i) Solve power flow with basic PV models and unbounded reactive power $Q_G$
ii) Project the unbounded solution back to the respective segments of the disjunctive function from (10) using predefined heuristics (usually based on the respective bus voltage sensitivities [27]), and keep re-solving the power flow in the outer loop until all of the devices fall on the respective segment of the VR characteristic.

From the perspective of numerical methods, this outer-loop approach represents a fixed-point iteration loop around the power flow NR solver, which can converge linearly but is also known to have serious drawbacks, such as oscillations and divergence. A more detailed discussion on this approach and its drawbacks can be found in [27]-[28].

- *Implicit continuous VR modeling*

Recent advances include continuous models that characterize the disjunctive VR behavior in terms of a steep sigmoid [28] or complementarity functions [11] and as such can be implicitly included within the NR power flow solver by replacing the $F_G$ from (9). Moreover, inspired by the steep nonlinearities that can be found in integrated circuit simulations, the circuit simulation heuristics developed solely for handling the steep VR nonlinearities [11] demonstrated extremely promising results [28] in terms of stability and efficiency over existing state-of-the-art.

Finally, independently of the method used to incorporate the realistic operational bounds on VR devices, the additional constraints only increase the likelihood of power flow problem infeasibility due to possible reactive power deficiencies in the system. A framework that can efficiently identify and quantify these deficiencies is therefore important.

### III. ADJOINT POWER FLOW NETWORK MODELS

The adjoint network concept introduced in [21]-[22] is a well-studied and understood concept that has been used for various circuit analyses, most notably noise analysis in SPICE [9]. It is derived from Tellegen's Theorem [3]. In the first part of this section, we provide a brief introduction to the adjoint network concept and apply it to the linear network elements of the power flow problem defined in terms of current and voltage state variables. Notably, an adjoint circuit methodology does not exist at present for nonlinear steady-state elements defined at fixed frequency (models that constrain the complex power), hence we further extend the adjoint network theory to derive the adjoint network models of nonlinear elements within the power flow (I-V) formulation.

#### A. Adjoint equivalent of linear power flow network elements

Consider a linear time-invariant network $\mathcal{N}$ and its topologically equivalent adjoint $\widehat{\mathcal{N}}$, where $\tilde{I}, \tilde{V}, \tilde{\mathfrak{T}}$ and $\tilde{\lambda}$ represent the network and adjoint branch current and voltage phasors respectively. From Tellegen's Theorem we can write the following relationship that has to be satisfied [3],[21]-[22]:

$$\tilde{\lambda}^H \tilde{I} - \tilde{\mathfrak{T}}^H \tilde{V} = 0 \tag{11}$$

Next, if the circuit equations of network $\mathcal{N}$ have a form of:

$$\tilde{I} = Y\tilde{V} \tag{12}$$

By substituting (12) in (11) we can obtain:

$$\tilde{\lambda}^H Y \tilde{V} - \tilde{\mathfrak{T}}^H \tilde{V} = 0 \tag{13}$$

Hence in order for Tellegen's Theorem to remain satisfied, the adjoint current $\tilde{\mathfrak{T}}$ that further defines the transformation from network $\mathcal{N}$ to its adjoint $\widehat{\mathcal{N}}$ has to be equivalent to:

$$\tilde{\mathfrak{T}} = Y^H \tilde{\lambda} \tag{14}$$

As it can be seen from (14), the linear sensitivity (admittance) matrix of the adjoint circuit corresponds to the Hermitian of the original admittance matrix. Furthermore, since independent voltage and current sources are constant, their sensitivities are zero and, therefore, represent an open and short, respectively, in the adjoint domain [21]. For instance, an independent voltage source that models the slack bus generator [25] is further represented by a zero-voltage source in the adjoint network. In the following, the generalized mapping of linear circuit elements that represent the building blocks of linear power flow models (transmission line, shunt, slack bus, transformer, etc.) are presented in Table 1. More detailed derivation of adjoint network models can be found in [11].

TABLE 1: MAPPING THE LINEAR ELEMENTS TO ADJOINT DOMAIN

| Powerflow network | | Adjoint network |
|---|---|---|
| Independent current source | → | open |
| Independent voltage source | → | short |
| Capacitor | → | Inductor |
| Inductor | → | Capacitor |
| Conductance | → | Conductance |

For the given series admittance and shunt susceptance of a transmission line model, we obtain its adjoint power flow equations by using the relation defined in (14). The complex governing equation for the series part is given by Ohm's Law that relates the complex adjoint circuit current ($\tilde{\mathfrak{T}}_{km} = \mathfrak{T}_{km}^R + j\mathfrak{T}_{km}^I$) and voltage ($\tilde{\lambda}_{km} = \lambda_{km}^R + j\lambda_{km}^I$) as:

$$\tilde{\mathfrak{T}}_{km} = (G_L - jB_L)\tilde{\lambda}_{km} \tag{15}$$

The corresponding real and imaginary adjoint currents of series elements are further obtained by splitting (15):

$$\mathfrak{T}_{km}^R = G_L \lambda_{km}^R + B_L \lambda_{km}^I \tag{16}$$

$$\mathfrak{T}_{km}^I = G_L \lambda_{km}^I - B_L \lambda_{km}^R \tag{17}$$

As in the case of mapping the power flow circuit, the terms where the adjoint current is proportional to the adjoint voltage drop across the line are modeled by conductance, while the other terms proportional to the voltage drop of the other adjoint circuit represent voltage-controlled current sources. Similarly, the complex current flowing through the shunt is given by (18) and once combined with the series elements from (16)-(17) correspond to the circuit in Fig. 5:

$$\mathfrak{T}_{k,sh}^R + j\mathfrak{T}_{k,sh}^I = \frac{B_{sh}}{2}\lambda_k^I - j\frac{B_{sh}}{2}\lambda_k^R \tag{18}$$

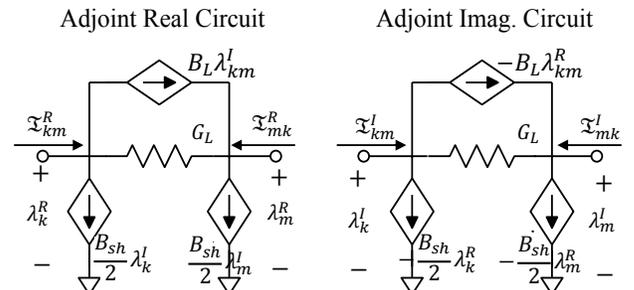

Fig. 5. Adjoint power flow split-circuit of a transmission line π-model.

#### B. Adjoint equivalent of nonlinear power flow models

Power flow analysis with the traditional constant power macromodels (e.g., PQ load, PV generator) can be classified as a nonlinear AC steady-state analysis for a fixed frequency point. This is not well explored in the circuit simulation field due to the lack of models that exhibit such behavior in integrated circuits. Namely, existing circuit theory recognizes two types of AC steady-state response modeling and analyses. The first, linear AC analysis, can be fully characterized by an RLC network whose response remains of the same harmonic frequency as its excitation and only changes in magnitude and phase. The second, nonlinear AC analysis [37], recognizes models that introduce distortion to the network response and hence its output is not of the same harmonic frequency as the network excitation. Power flow analysis can be seen as an introduction of the nonlinearities in constraining the powers of a linear AC network response. Therefore, in order to derive the adjoint split-circuit models of nonlinear power flow elements (PV generator, PQ load), we generalize the linear adjoint circuit methodology discussed in III.A.

Consider the split-circuit equations that govern the nonlinear power flow elements, expressed at the solution in terms of their first order sensitivities $\mathcal{J}(V)$ (small signal model [10]):

$$I_{NL}(V) = \mathcal{J}(V) V + \alpha \quad (19)$$

where $I_{NL}$, $V$ and $\alpha$ represent the nonlinear split-circuit currents $(I_R, I_I)$, state variables, and independent sources, respectively.

Since the independent sources $\alpha$ do not contribute to the adjoint circuit, they can be omitted in further derivations without loss of generality. Next, we substitute (19) into the generalized relationship obtained from Tellegen's Theorem in (11) to obtain the expression for the nonlinear adjoint current that further defines the transformation from power flow to the adjoint network:

$$\mathfrak{T}_{NL} = \mathcal{J}(V)^T \lambda \quad (20)$$

Note that if the sensitivity matrix $\mathcal{J}(V)^T$ is linear, (20) becomes equivalent to the split-circuit form of (14). Otherwise, the nonlinear elements from power flow also introduce nonlinearities within the adjoint power flow circuit. The more detailed derivation of generalized adjoint network models can be found in [11].

*1) Adjoint model of a PQ load*

We start the derivation of the adjoint split-circuit of a PQ load by rewriting the power flow split-circuit governing equations from (7)-(8) in the form given in (19).

$$\begin{bmatrix} I_{RL} \\ I_{IL} \end{bmatrix} = \begin{bmatrix} \frac{\partial I_{RL}}{\partial V_{RL}} & \frac{\partial I_{RL}}{\partial V_{IL}} \\ \frac{\partial I_{IL}}{\partial V_{RL}} & \frac{\partial I_{IL}}{\partial V_{IL}} \end{bmatrix} \begin{bmatrix} V_{RL} \\ V_{IL} \end{bmatrix} + \begin{bmatrix} \alpha_{RL} \\ \alpha_{IL} \end{bmatrix} \quad (21)$$

The nonlinear adjoint circuit equations that define the PQ load can be further obtained from (20) as:

$$\begin{bmatrix} \mathfrak{T}_{RL} \\ \mathfrak{T}_{IL} \end{bmatrix} = \begin{bmatrix} \frac{\partial I_{RL}}{\partial V_{RL}} & \frac{\partial I_{IL}}{\partial V_{RL}} \\ \frac{\partial I_{RL}}{\partial V_{IL}} & \frac{\partial I_{IL}}{\partial V_{IL}} \end{bmatrix} \begin{bmatrix} \lambda_{RL} \\ \lambda_{IL} \end{bmatrix} \quad (22)$$

Further, since the current sensitivities from (22) represent nonlinear functions of real and imaginary power flow voltages, we further linearize the nonlinear adjoint load currents using the first order Taylor expansion. We map the equations to an equivalent circuit, where the current terms that are proportional to the adjoint voltage across the load terminals represent conductances, and the terms related to the voltages in the opposite sub-circuit define controlled-current sources. Historical terms known from the previous iteration are given by independent current sources.

*2) Adjoint model of a PV generator*

The governing power flow split-circuit equations of a PV generator can be expressed in the form defined by (19) as:

$$\begin{bmatrix} I_{RG} \\ I_{IG} \\ 0 \end{bmatrix} = \begin{bmatrix} \frac{\partial I_{RG}}{\partial V_{RG}} & \frac{\partial I_{RG}}{\partial V_{IG}} & \frac{\partial I_{RG}}{\partial Q_G} \\ \frac{\partial I_{IG}}{\partial V_{RG}} & \frac{\partial I_{IG}}{\partial V_{IG}} & \frac{\partial I_{IG}}{\partial Q_G} \\ \frac{\partial F_G}{\partial V_{RG}} & \frac{\partial F_G}{\partial V_{IG}} & \frac{\partial F_G}{\partial Q_G} \end{bmatrix} \begin{bmatrix} V_{RG} \\ V_{IG} \\ Q_G \end{bmatrix} + \begin{bmatrix} \alpha_{RG} \\ \alpha_{IG} \\ f_G \end{bmatrix} \quad (23)$$

The nonlinear adjoint circuit of a PV generator is then obtained by applying the transformation from (20):

$$\begin{bmatrix} \mathfrak{T}_{RG} \\ \mathfrak{T}_{IG} \\ 0 \end{bmatrix} = \begin{bmatrix} \frac{\partial I_{RG}}{\partial V_{RG}} & \frac{\partial I_{IG}}{\partial V_{RG}} & \frac{\partial F_G}{\partial V_{RG}} \\ \frac{\partial I_{RG}}{\partial V_{IG}} & \frac{\partial I_{IG}}{\partial V_{IG}} & \frac{\partial F_G}{\partial V_{IG}} \\ \frac{\partial I_{RG}}{\partial Q_G} & \frac{\partial I_{IG}}{\partial Q_G} & \frac{\partial F_G}{\partial Q_G} \end{bmatrix} \begin{bmatrix} \lambda_{RG} \\ \lambda_{IG} \\ \lambda_V \end{bmatrix} \quad (24)$$

As in the case of the nonlinear adjoint PQ load, the first two equations from (24) represent the nonlinear real and imaginary adjoint generator currents. These can be linearized by a first order Taylor expansion to define the adjoint split-circuit. The third equation represents the adjoint equivalent of the voltage magnitude constraint and can be simplified for the basic (without VR control incorporated) PV model as:

$$V_{IG} \lambda_{RG} - V_{RG} \lambda_{IG} = 0 \quad (25)$$

It can be shown that the relationship between power flow and adjoint bus voltages from (25) constrains the PV bus voltage angle of the adjoint circuit to be equal to the respective voltage angle of the power flow circuit. Equation (25) is then linearized and stamped for the $\lambda_V$ adjoint variable as in the case of the voltage magnitude constraint.

Finally, with VR control characteristics included, adjoint model of a PV generator remains as given by (24), with the key differences found in the sensitivities of the $F_G$ equation. Namely, in the 'outer loop' approach the adjoint network is included within the loop and the $F_G$ sensitivities correspond to the current segment of the disjunctive function, while in the implicit VR model they simply represent the sensitivities of the continuous model governing equation.

## IV. LOCATING AND EVALUATING POWER FLOW INFEASIBILITY WITH ADJOINT NETWORK

With the relationship between the power flow and adjoint networks established, in this section we demonstrate the use of adjoint network theory to represent first-order optimality conditions for the grid optimization problems. Specifically, we will show how the adjoint network is used to solve for and localize power flow infeasibilities. To do so, we couple the power flow network with its corresponding adjoint network through the use of *feasibility current sources* ($I_F$) and show that if a power flow solution does exist, the adjoint network converges to a solution that has zero voltage at all nodes. However, when the system is infeasible, the *feasibility current sources* will optimally "pick up the slack" and prevent KCL violations at the corresponding nodes by using it to set an operating point of the adjoint circuit. The non-zero voltages in the adjoint network now indicate the nodes that contribute most toward the infeasibility in the grid. The *feasibility current sources* are added to each node in the power flow circuit are shown in Fig. 6, whereas the adjoint circuit for the corresponding sources are derived per the methodology in Section III.

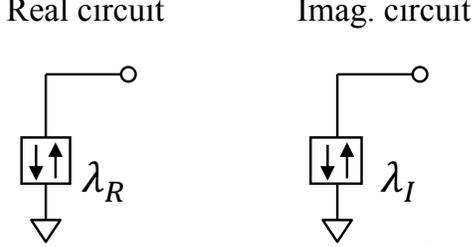

Fig. 6. Infeasibility current sources added to the power flow circuit.

Next, we demonstrate how the solution of the coupled power flow and adjoint networks with feasibility current sources corresponds to the solution of the optimization problem of minimizing the L2-norm of these current sources subject to the power flow network constraints.

First, consider the generalized form of equations that represent the power flow and its adjoint networks coupled through the feasibility currents, namely (12), (19)-(20):

$$\begin{bmatrix} Y_{GB} + \mathcal{J}(V^k) & -\hat{\mathbf{1}} \\ \frac{\partial \mathfrak{T}_{NL}(V^k, \lambda^k)}{\partial V} & Y_{GB}^T + \mathcal{J}^T(V^k) \end{bmatrix} \begin{bmatrix} V \\ \lambda \end{bmatrix} = -\begin{bmatrix} \alpha \\ \beta \end{bmatrix} \quad (26)$$

where $Y_{GB}$ represents the linear admittance matrix from (12), defined for the split-circuit models and $\hat{\mathbf{1}}$ is a degenerate identity matrix with zero diagonal entries corresponding to the indices of voltage magnitude constraints.

Next, in order to show how the set of circuit equations from (26), relates to the necessary optimality conditions of an optimization problem, we define the following program:

$$\min_{I_F} \frac{1}{2} \|I_F\|_2^2 \quad (27)$$

subject to power flow constraints with additional feasibility currents:

$$Y_{GB}V + I_{NL}(V) = I_F \quad (28)$$

To find the optimality conditions of the optimization problem defined by (27)-(28), we define the Lagrangian function as:

$$\mathcal{L}(V, I_F, \lambda) = \frac{1}{2}\|I_F\|^2 + \lambda^T(Y_{GB}V + I_{NL}(V) - I_F) \quad (29)$$

The necessary KKT optimality conditions are further obtained by differentiating (29) with respect to power flow and adjoint variables, as well as newly introduced current ($I_F$) variables:

$$\frac{\partial \mathcal{L}}{\partial V} \to [Y_{GB}^T + \mathcal{J}^T(V)]\lambda = 0 \quad (30)$$

$$\frac{\partial \mathcal{L}}{\partial I_F} \to I_F = \lambda \quad (31)$$

$$\frac{\partial \mathcal{L}}{\partial \lambda} \to [Y_{GB} + \mathcal{J}(V)]V - I_F + \alpha = \mathbf{0} \quad (32)$$

After linearizing (30) and eliminating the ($I_F$) variables by substituting (31) in (32), we end up with the system of equations from (26). From this we postulate the following theorem.

**Theorem 1**. *Let $\mathcal{N}$ and its topologically equivalent adjoint $\widetilde{\mathcal{N}}$ represent the power flow and adjoint networks respectively that are further coupled through infeasibility current sources connected to every node of the power flow network $\mathcal{N}$. An operating point of such jointly coupled system model then corresponds to a power flow solution that can further capture its infeasibilities. Moreover, if the sufficient optimality condition* (33) *is met, an operating point also provides with a minimal current flow between the power flow circuit and its adjoint; i.e. the infeasibility currents are minimized.*

*Proof.* Following from (26) and (30)-(32), the governing equations of the proposed problem represent the necessary KKT optimality conditions of the optimization problem given by (27)-(28). Hence, any network operating point of the jointly coupled power flow and its adjoint network models represents an optimal solution candidate to (27)-(28) and as such locates the possible problem infeasibilities.

i) Importantly, if an obtained operating point indicates the trivial adjoint network response (zero adjoint voltages), the power flow problem is feasible and a network operating point also represents an optimal solution to (27)-(28).

ii) If, however, the adjoint network response is not trivial, an obtained operating point indicates the power flow infeasibilities. Therefore, due to the constant power-based models that introduce the nonlinearities within the formulation, a network operating point represents an optimal solution if the second order optimality sufficient condition [29] holds:

$$\tau^T \nabla_{xx}^2 \mathcal{L}(V^*, \lambda^*)\tau > 0 \quad \forall (\tau \neq \mathbf{0}) \in T_{V^*} \quad (33)$$

where $V^*$ is an obtained network operating point and $T_{V^*}$ represents a null space of the power flow network sensitivity matrix, i.e. $\text{Null}[Y_{GB} + \mathcal{J}(V^*) - \hat{\mathbf{1}}]$.

Lastly, if a case of any of the two scenarios (ii) or (iii), the network operating point obtained from a jointly solved power flow and its adjoint network, in addition to locating the source of infeasibility for the simulation model, optimally allocates (minimizes) the respective current violations. ∎

From the perspective of the optimization problem, the infeasibility current sources do not have to be added to the voltage magnitude constraints given by (8), since by Ohm's Law and KCL there is always a current that can be injected into a node of the power flow circuit that prevents the system solution from being infeasible. *This further eliminates the need for multi-objective optimization and additional weighting factors that have to be assigned in order to obtain a physically meaningful optimal solution*, as required by other proposed approaches to determine the power flow infeasibility [16].

V. BUILDING AND SOLVING AN EQUIVALENT SPLIT-CIRCUIT

*A. Generalized power system problem*

A current-voltage power flow formulation represented by an equivalent split-circuit was demonstrated to provide a generalized power system simulation framework [4]-[7], [30]-[31]. As it is shown in the previous sections, each of the power system device models (PV generator, PQ load, etc.) can be further defined within the adjoint (dual) domain. The complete split-circuit representation is then obtained by hierarchically combining (connecting) the derived power flow and adjoint power flow circuit models, as defined by the grid (network) topology. Importantly, the hierarchical building of the circuit representation corresponds to a modular construction of the Jacobian/Hessian matrix and constant vector that defines the Newton-Raphson (NR) values during the iteration process.

Coupling the power flow with its adjoint circuit corresponds to solving an optimization problem whose objective is specified by the type of coupling between the two circuits. *This further defines a new class of optimization problems, Equivalent Circuit Programming (ECP), whose constraints can be expressed in terms of equivalent circuit equations, and whose solutions can therefore be obtained by solving circuit simulation problems.* Adding the adjoint sources to the power flow circuit to capture infeasibility can be done in the beginning of the simulation, where simulating the circuit corresponds to solving an optimization problem, or during the power flow simulation, when the iterative simulation method starts diverging, thereby indicating possible infeasibility.

### B. Generalized solution of an equivalent split-circuit

Once the complete equivalent split-circuit is built, its set of governing circuit equations correspond to the linearized set of equations that are updated at each step of NR. In equivalent circuit approach to NR, only circuit elements (Jacobian/Hessian terms) that are dependent on the values from the previous iteration are recomputed, while the linear parts are only computed once at the beginning of the simulation. This approach was shown to represent an extremely efficient formulation and solution method for solving the nonlinear circuit problems [3],[10]. The main difference between the circuit simulation and traditional NR method, however, is the circuit formalism obtained from the circuit representation of the problem. This provides important information that allows for developing efficient heuristics for limiting the Newton step, thereby ensuring robust and efficient convergence properties [4]-[7],[10] as is the case in the circuit simulation field. Embedding the physical characteristics of the problem in the NR-step control methods obtained from the circuit perspective of the power flow problem is discussed in Section VI.

Initialization of the power flow split-circuit is well defined by the power flow problem as it is specified in [4]-[7]. When initializing the adjoint power flow circuit, we need to consider that the adjoint voltages correspond to the magnitude of the respective infeasibility currents in the power flow analysis. Hence, we initialize them to a small constant value, such as the NR tolerance used for the convergence criterion. After the circuit initialization and first iteration, the linearized circuit elements are updated as discussed in the following section.

## VI. EXTENDING THE CIRCUIT SIMULATION METHODS

The circuit formalism was demonstrated to provide understanding of the characteristics of each power flow state variable and its sensitivities directly from first principles. As it was shown in [5]-[7], during the solution process of a power flow problem, a large NR step may lead the solution trajectory out of a well-defined solution space and result in either divergence or convergence to a non-physical solution. It is, therefore, crucial to limit the NR step before it makes an invalid step out of the solution space. Furthermore, limiting methods may fail to converge for large-scale ill-conditioned test cases solved from an arbitrary initial guess. Hence, the use of homotopy methods, such as "Tx-stepping" in [6], and other developed homotopy methods [8] can be crucial to ensure convergence. Importantly, the nonlinearities of the adjoint split-circuit resemble the ones that are robustly handled within the power flow problem, while the feasibility of the simulation problem is ensured. Thus, in this section we extend the circuit simulation limiting and homotopy methods to ensure robust convergence of any size power systems.

### A. Voltage limiting

Voltage Limiting was shown to be a simple and effective simulation technique that limits the absolute value of the step change that the real and imaginary voltage vectors are allowed to make during each NR iteration [4]-[7]. The power flow voltage step limiting technique is given in compact form as:

$$\xi_{BC,i} = \min\left[1, \text{sign}(\Delta B_{C,i}^k)\frac{\Delta B_{max}}{\Delta B_{C,i}^k}\right], \forall i \in [1, |\Delta B_C^k|] \quad (34)$$

where placeholders $C \in \{R, I\}$ and $B \in \{V, \lambda\}$ in $\Delta B_{C,i}^k$ represent the power flow and adjoint voltage NR steps, while $\Delta B_{max}$ is a maximum allowable step change.

Furthermore, the hard limits can be imposed to prevent the voltage variables to escape the physical solution space as shown in [23]. Lastly, the obtained limiting factors are used to limit the step change of power flow and adjoint power flow voltages as:

$$\widehat{V}_C^{k+1} = V_C^k + \xi_{VC} \odot \Delta V_C^k \quad (35)$$
$$\widehat{\lambda}_C^{k+1} = \lambda_C^k + \xi_{\lambda C} \odot \Delta \lambda_C^k \quad (36)$$

### B. Tx-Stepping homotopy method

To achieve robust convergence of large-scale power system simulation problem, we extend the recently introduced Tx-stepping method [6] to the simulation discussed in this paper.

The solution of the feasibility simulation problem (coupled power flow and adjoint power flow circuits), is obtained by embedding the homotopy factor $\mu \in [0,1]$ to linear series network elements and transformer model as shown in (37)-(39) and sequentially solving the relaxed feasibility problems while gradually decreasing the homotopy factor to zero. Namely, for the initial homotopy factor set to one, the circuit of the feasibility problem is virtually "shorted". Now, the power flow solution is feasible and driven by the generator voltages and the slack bus angle and can be trivially obtained. Gradual decreasing of the embedded homotopy factor $\mu$ to zero sequentially relaxes the feasibility circuit toward its original state, while using the solution from the previous sub-problem to initialize the circuit for the next homotopy decrement:

$$G_{km} + jB_{km} \rightarrow (\mu\Upsilon + 1)(G_{km} + jB_{km}) \quad (37)$$
$$t(\mu) \rightarrow t + (1-t)\mu \quad (38)$$
$$\theta_{ph}(\mu) \rightarrow (1-\mu)\theta_{ph} \quad (39)$$

where $\Upsilon$ represents an admittance scaling factor, $t$ is the transformer tap, and $\theta_{ph}$ is the phase shifting angle.

Most importantly, any homotopy-embedded model defined for the adjoint circuit has to be governed by transformations given in (14) and (20). Therefore, transformer and phase shifter parameters remain scaled by (38)-(39) within their respective adjoint models, while the homotopy admittance are conjugated.

## VII. Simulating and Locating Power Flow Infeasibility

The circuit element library for the introduced adjoint power flow models was built and incorporated within our prototype simulator Simulation with Unified Grid Analyses and Renewables (SUGAR). Since most standard commercial power grid simulation tools do not locate and quantify the power flow infeasibilities, in order to validate the proposed approach, we first examine that our solution exactly matches a standard power flow solution for a feasible case. Therefore, we tested our approach using three different categories of test cases available in the literature: 1. **Benchmark ill-conditioned test cases (case11 and case145)** [32]-[33], **2. European RTE and PEGASE test cases** [34], **3. Synthetic USA test cases** [35]. Each of the test cases is run on a machine with an Intel Core i7-6700 3.4GHz processor, and it is confirmed that the same solution is obtained between the proposed approach and the traditional power flow approach.

Notably, the addition of the adjoint power flow split-circuit increases the size of the simulation problem. Therefore, to study the effect on the simulation complexity, in Fig. 7 we compare the average runtime per iteration obtained from the simulation of a variety of test cases for regular power flow in SUGAR, regular power flow in a commercial tool, and adjoint/feasibility-based power flow in SUGAR. As expected, there is an increase in average time per iteration with respect to the power flow run in SUGAR without the adjoint (feasibility) option enabled. However, the increase in runtime is justified as regular power flow solvers would not converge for an infeasible test case, leaving the user to determine where the infeasibility stems from. With adjoint power flow in SUGAR, the location and amount of infeasibility is automatically computed. Most importantly, the simulation runtime is not significantly affected, which makes the proposed formulation a promising tool for future implementations in contingency simulations.

Lastly, we examine the power flow convergence characteristics and demonstrate how embedding domain-specific knowledge within a set of simulation algorithms used to control the NR-step can improve efficiency and robustness. A comparison of iteration count between SUGAR and the four following formulations

i. Power mismatch in polar coordinates (Polar PQV)
ii. Power mismatch in rectangular coordinates (Rect. PQV)
iii. Current Injection in polar coordinates (Polar I-V)
iv. Current Injection in rectangular coordinates (Rect. I-V)

within the MATPOWER [33] simulator is provided in Table II.

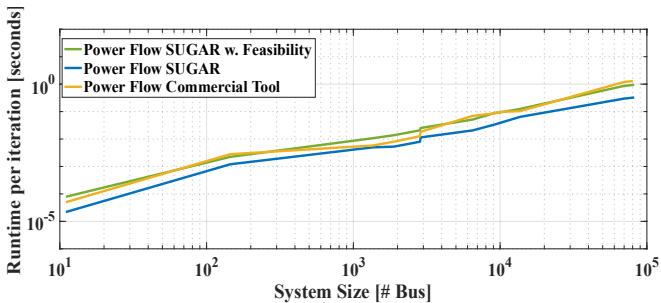

Fig. 7. Average runtime per iteration comparison

*Table 2. Iteration count comparison between the existing state-of-art power flow formulations and SUGAR*

| Formulation | Polar PQV | | Rect. PQV | | Polar I-V | | Rect. I-V | | **SUGAR*** | |
|---|---|---|---|---|---|---|---|---|---|---|
| case | Flat | Input | Flat | Input | Flat | Input | Flat | Input | Flat | Input |
| 1354pegase | 5 | 4 | 6 | 4 | ∞ | 4 | ∞ | 4 | **4** | **1** |
| 1888rte | ∞ | 2 | ∞ | 2 | ∞ | 2 | ∞ | 2 | **5** | **1** |
| 1951rte | ∞ | 3 | ∞ | 3 | ∞ | 3 | ∞ | 3 | **5** | **1** |
| 2383wp | 5 | 6 | 5 | ∞ | 6 | 5 | 6 | 5 | **3** | **2** |
| 2736sp | 6 | 4 | ∞ | 4 | ∞ | 4 | ∞ | 4 | **3** | **2** |
| 2746wp | 6 | 5 | ∞ | 5 | ∞ | 4 | ∞ | 4 | **3** | **1** |
| 2848rte | 10 | 3 | ∞ | 3 | ∞ | 3 | ∞ | 3 | **2** | **1** |
| 2869pegase | 5 | 7 | 8 | 6 | ∞ | 4 | ∞ | 4 | **4** | **1** |
| 3012wp | ∞ | 3 | ∞ | 3 | ∞ | 3 | ∞ | 3 | **3** | **1** |
| 3120sp | 6 | 6 | ∞ | ∞ | 8 | 8 | ∞ | ∞ | **3** | **2** |
| 3375wp | ∞ | 2 | ∞ | 2 | ∞ | 2 | 14 | 2 | **4** | **1** |
| 6468rte | ∞ | 3 | ∞ | 3 | ∞ | 3 | ∞ | 3 | **3** | **1** |
| 6515rte | ∞ | 3 | ∞ | 3 | ∞ | 3 | ∞ | 3 | **5** | **1** |
| 9241pegase | 7 | 6 | ∞ | 6 | ∞ | 4 | ∞ | 4 | **4** | **1** |
| ACTIVSg10k | ∞ | 5 | ∞ | 5 | ∞ | 5 | ∞ | 5 | **5** | **2** |
| 13659pegase | ∞ | 6 | ∞ | 6 | ∞ | 5 | ∞ | 5 | **5** | **2** |
| ACTIVSg25k | 6 | 5 | ∞ | 5 | ∞ | 4 | ∞ | 4 | **5** | **1** |
| ACTIVSg70k | ∞ | 6 | ∞ | 8 | ∞ | 5 | ∞ | 5 | **7** | **2** |

∞-indicates the divergence of the simulation
*Iteration counts within SUGAR were identical for both Power Flow and Feasibility analysis

In referring to the formulations solved within MATPOWER *only*, the initial starting point played a significant role in the convergence process. As can be seen, most of the cases converged when initialized with a good starting point that usually represents an operating point close to the actual solution. Moreover, the lack of robustness when a good initial guess is not provided is particularly emphasized as the sizes of the cases increase. By further comparing the examined formulations solved without utilizing any of the domain specific knowledge, the overall performance of Polar PQV formulation with provably [38] positive definite Jacobian matrix is slightly better in reference to the other three formulations. The positive definite Jacobian, however, does not guarantee the convergence, and therefore, it can be seen that there are some cases that converge with other formulations while not converging with a Polar PQV

Utilizing the physical characteristics of the problem and further embedding the domain knowledge within the set of algorithms used to control the NR step, as done in SPICE, resulted in a significant improvement over the traditionally implemented formulations solved within the MATPOWER simulator. This is particularly evident when a good initial start is not provided. Most importantly, the iteration count profile of the examined test cases does not change with the introduction of the adjoint network.

Next, we apply the proposed power flow feasibility framework to examine and locate infeasibilities that may arise due to the operation at the edge of voltage collapse or a contingency.

### 1. Synthetic USA grid during a loading factor change

Particularly in large-scale simulations of the extreme planning power flow cases, the initial power flow starting point may not be sufficiently close to the actual operating point, which can result in simulation divergence [7]. More importantly, if the power flow case diverges, it is generally not known whether the divergence is caused by a lack of simulation robustness or true problem infeasibility. Therefore, to demonstrate the robustness of the circuit simulation methods, we consider the 80k-bus synthetic USA test case [35] and perform power flow simulations for three different loading factors in SUGAR and a commercial simulator initialized from the solution in the input file.

TABLE 3: SYNTHETIC USA CONVERGENCE PROFILE COMPARISON

| Simulator | Loading Factor | | |
|---|---|---|---|
| | 0.8 | 1 | 1.1 |
| Commercial tool | Diverge | Converge | Diverge |
| SUGAR (feasibility OFF) | Converge (**Feasible**) | Converge | Diverge |
| SUGAR (feasibility ON) | Converge (**Feasible**) | Converge | Converge (**Infeasible**) |

As it can be seen from Table 2, both the commercial tool and SUGAR converge for the base loading factor. The commercial tool fails to converge once the loading factor is changed and indicates to the user that the system has experienced blackout conditions. However, the power flow case is truly infeasible only for the increased loading factor, while the feasible solution for the decreased loading factors exists and can be obtained within SUGAR. Lastly, the divergence of SUGAR without the feasibility option corresponds to true infeasibility that can be computed only once the feasibility option is enabled.

### 2. *Existing state-of-art optimization algorithms*

The first step in demonstrating the efficiency of the SUGAR framework in solving the power flow feasibility problem is to compare it with the existing state-of-the-art algorithms implemented within optimization toolboxes. Therefore, in order to have a fair comparison between the two, and considering the efficiency of the circuit simulation approach for building the Hessian and Jacobian matrices, we have developed a prototype circuit simulator in MATLAB, where the Gradient and Hessian information are built in the same way as in SUGAR, and the *only difference represents the respective set of existing optimization algorithms applied to control the NR step size*. For this comparison, we used the optimization algorithms implemented within the FMINCON toolbox in MATLAB. The runtime comparisons are performed using the MATLAB SUGAR prototype simulator on a MacBook Pro 2.9 GHz Intel Core i7 for the following five test cases: (1) 3375wp, (2) 9241pegase, (3) ACTIVSg10k, (4) ACTIVSg25k and (5) ACTIVSg70k [35], that are further solved for three operating conditions (nominal loading conditions as defined in the file, generation/load increase of 25%, and generation/load decrease of 25%). The runtime comparisons for initializing with the input file operating point and a flat start are given in Fig. 8-Fig. 10. As can be seen from the respective figures:

i. It can be said that the two smaller-size test cases performed equally well with both SUGAR and traditional heuristics for all scenarios and initializations. This can be further explained by the initial starting points of the examined equality constrained optimization problems being in the vicinity of the respective optimal solutions.

ii. As the size of problem increases and the initial starting points move away from the optimization solutions (increased loading, flat start, etc.), the traditional heuristics start slowing down the convergence process due to the single step line-search methodology that limits all of the NR steps with a single constant factor [29].

iii. In contrast to the traditional optimization heuristics, however, the heuristics based on physical characteristics of the problem incorporated within the SUGAR solver significantly outperforms the traditional optimization heuristics particularly when it comes to larger scale test cases.

iv. Finally, the experiments performed also indicate the importance of a robust and efficient simulation and optimization framework when a good initialization is not known, such as for extreme contingency and other planning analyses. The preliminary results indicate a significant correlation between the "goodness" of the initialization and the traditional simulator efficiency, which is particularly highlighted as the size of cases increases.

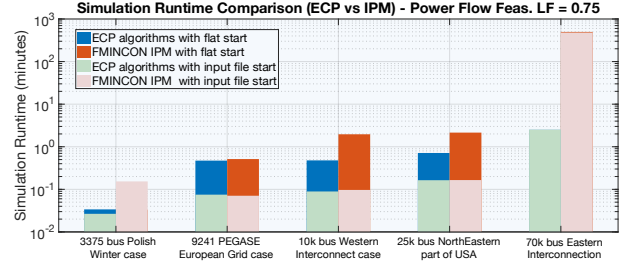

Fig. 8. Simulation Runtime comparison. SUGAR vs. traditional state-of-the-art algorithms for 'Off-peak' loading conditions.

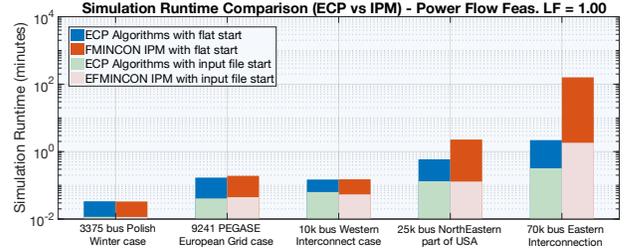

Fig. 9. Simulation Runtime comparison. SUGAR vs. traditional state-of-the-art algorithms for nominal loading conditions.

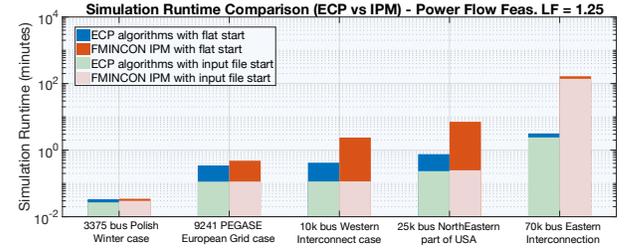

Fig. 10. Simulation Runtime comparison. SUGAR vs. traditional state-of-the-art algorithms for 'On-peak' loading conditions.

### 3. *Feasibility analysis of ill-conditioned 11 bus test case*

The authors in [32] have demonstrated that the 11-bus distribution test case is genuinely ill-conditioned beyond a maximum loading factor of 99.82%. Hence, numerical error or the choice of convergence criterion can cause the difference between infeasibility (divergence of the numerical algorithm) or convergence to the operating solution.

In this study, the power flow feasibility analysis is solved for slight loading factor increments to locate and examine the appearance and evolution of infeasible regions within the test case. The simulation results representing the network topology for four different loading factors (three of which are provably infeasible) are presented in Fig. 11. Referring to Fig. 11, after the known point of collapse is reached, the system first becomes infeasible (indicated by the heatmap around the infeasible bus) furthest from the slack generator (bus 11). As the loading factor keeps increasing, the infeasibility, which represents the amount of additional current needed to prevent the violation of KCL at each bus, evolves throughout the system.

To compare the proposed Power Flow Feasibility Analysis (PFFA) solved using the circuit simulation techniques and MATLAB 'FMINCON' solver with an existing traditional formulation that minimizes infeasible real and reactive power injection within the power-mismatch formulation, we implemented the formulation from [16] in 'FMINCON' with the MATLAB optimization toolbox. The iteration count and total infeasible p.u. real and reactive powers ($P_{INF}$ and $Q_{INF}$) for two loading factors are shown in Table 3.

As expected, both formulations were able to converge to the same optimal infeasibility values from a flat start. The proposed PFFA that was solved as a circuit simulation problem, however, converged to the optimal solution much faster, especially near the system collapse point where the power flow Jacobian is singular. This can be attributed to the efficient circuit simulation limiting heuristics that we apply in our simulation framework [4]-[7] for faster convergence.

TABLE 4. RESULT COMPARISON FOR ILL-CONDITIONED 11 BUS CASE

| Formulation | Loading factor: 1.0000 | | | Loading factor: 1.1000 | | |
|---|---|---|---|---|---|---|
| | Iter. | $P_{INF}$ [p.u.] | $Q_{INF}$ [p.u.] | Iter. | $P_{INF}$ [p.u.] | $Q_{INF}$ [p.u.] |
| PFFA + ckt. techniques | 5 | 3.18E-4 | 2.59E-4 | 4 | 0.040 | 0.032 |
| PFFA with FMINCON | 45 | 3.18E-4 | 2.59E-4 | 34 | 0.040 | 0.032 |
| PQ Formulation [16] | 73 | 3.18E-4 | 2.59E-4 | 29 | 0.040 | 0.032 |

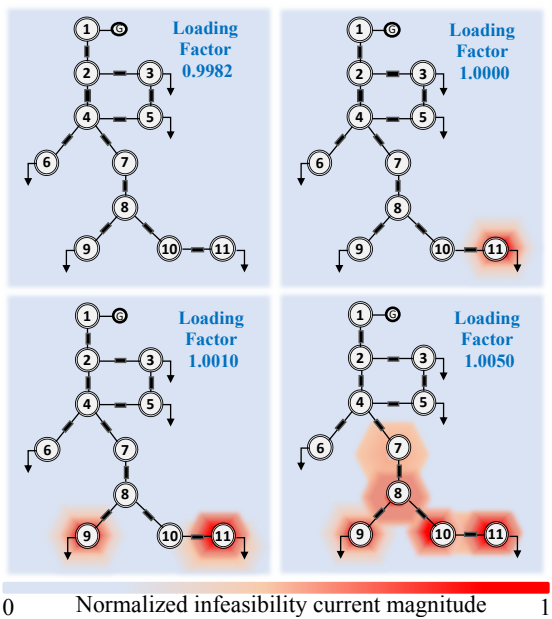

Fig. 11. Evaluating the feasibility of ill-conditioned 11 bus test case. The magnitude of infeasibility current is normalized with respect to the highest one encountered throughout the simulation of all four cases.

### 4. Remedial action for located infeasibility

Divergence of the power flow simulation during an N-1 analysis does not provide complete information about the analyzed grid. In this example, we study a test case of a real power system with over 5k buses that collapses under a contingency. Solving the case using the proposed method provided information about the localized area that caused the infeasibility due to a reactive power deficiency, as shown in Fig. 12. By activating a continuous FACTS device near the infeasible region, we were able to restore the system feasibility.

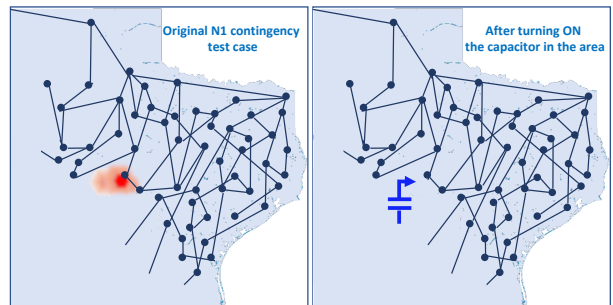

Fig. 12. Detecting and correcting the infeasibility on a real-life contingency case. Note that the network connectivity doesn't represent the true connectivity of the examined case.

### 5. Synthetic USA grid test case during N-1 contingency

The scalability of the proposed PFFA formulation is further tested by analyzing the feasibility of a test case representing the entire US grid, consisting of synthetic versions of the Eastern and Western Interconnections (WECC) as well as the ERCOT grid [35], during an N-1 contingency. The N-1 contingency we applied represented disconnecting the branch between buses 23510 (SENECA 71) and 23515 (SENECA76) within the Oconee Nuclear station, near Seneca, SC. PFFA simulation converged in 10 iterations and the results indicate that this contingency represents an infeasible system, with the local area of infeasibility shown in Fig. 13.

After analyzing the affected infeasible area and replacing the fixed shunt capacitor connected at the most infeasible bus (SENECA 7.1) with a continuous shunt device, the system becomes feasible again. Most importantly, as in the previous real-life contingency test case, the detected infeasibilities were local and mostly due to reactive power deficiency. In this case, by replacing the fixed shunt at bus number 23510 with continuous shunt, we were able to restore the test case to a feasible state. It should be noted that actual U.S. Eastern Interconnection testcases were found to be N-1 secure [6].

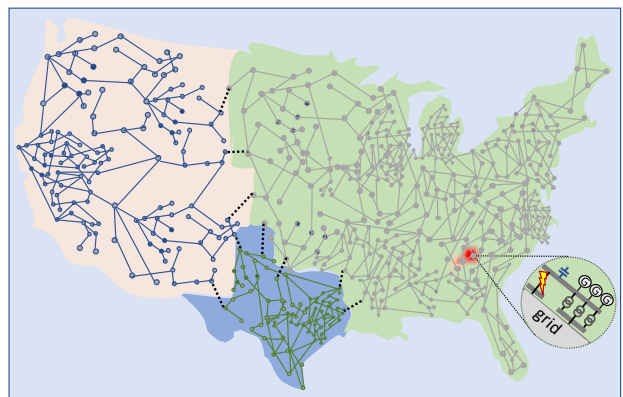

Fig. 13. Detecting infeasibility due to the contingency in synthetic test case representation of USA grid. Note that replacing the fixed shunt capacitor at bus 23515 with a variable capacitor restores the feasibility of the power flow.

Lastly, as shown by the simulation results, the placement of infeasibility current sources within the power flow circuit can inform various corrective actions or planning decisions. For instance, the 11-bus test case presents an application toward *optimal load shedding*, while the real-life contingency test case *optimally indicated the reactive power compensating device* that has to be activated in order to restore feasibility of the network. Hence the respective placement of infeasibility current sources to all load buses and existing variable shunt devices will provide all the necessary information needed to analyze the feasibility of the simulation. Importantly, placing the infeasibility sources at nodes of critical infrastructure within the grid model can allow optimal planning for a new corrective device that would *ensure N-1 contingency criteria required by NERC are met* [36].

## VIII. Conclusions

In this paper we introduced the framework for evaluating the feasibility of power flow cases as an extension of the recently introduced circuit theoretic approach for simulating the power flow problem. The infeasibility current sources and adjoint power flow network models are derived, and once coupled to the power flow circuit are shown to optimally prevent violation of KCL that arise for infeasible systems as a consequence of constant power models. Subsequently, recently proposed circuit simulation techniques are extended to ensure robust convergence properties of the adjoint power flow circuit. Finally, the proposed framework was demonstrated to provide an efficient methodology for locating and evaluating power flow infeasibilities. The authors believe the proposed framework has a number of practical applications within transmission planning and operations. Localized and quantified infeasibilities can help engineers pinpoint collapsed load pockets in large networks, design preventive and corrective actions to resolve voltage collapse, and identify regions of reactive and/or real power deficiencies when planning for high renewable penetration [39]

## IX. References

[1] EPRI. Application of advanced data processing, mathematical techniques and computing technologies in control centers: Enhancing speed and robustness of power flow computation, Dec. 2012. Technical Update.
[2] P. Kundur, N.J. Balu, and M.G. Lauby, "Power system stability and control". New York: McGraw-Hill, 1994, vol.7.
[3] W. J. McCalla, "Fundamentals of Computer-Aided Circuit Simulation", Kluwer Academic Publishers, Boston, 1988.
[4] M. Jereminov, et.al., "Improving Robustness and Modeling Generality for Power Flow Analysis," IEEE *T&D, 2016 IEEE PES*.
[5] A. Pandey, et. al., "Improving Power Flow Robustness via Circuit Simulation Methods," *IEEE PES General Meeting*, Chicago, 2017.
[6] A. Pandey, M. Jereminov, M. Wagner, G. Hug, L. Pileggi, "Robust Convergence of Power Flow using Tx Stepping Method with Equivalent Circuit Formulation" XX (PSCC), Dublin, Ireland, 2018.
[7] A. Pandey, M. Jereminov, M. Wagner, D. M. Bromberg, G. Hug, L. Pileggi, "Robust Power Flow and Three Phase Power Flow Analyses", IEEE Trans. on Power Systems. DOI: 10.1109/TPWRS.2018.2863042.
[8] M. Jereminov, A. Terzakis, M. Wagner, A. Pandey, L. Pileggi, "Robust and Efficient Power Flow Convergence with G-min Stepping Homotopy Method," in Proc. IEEE Conference on Environment, Electrical Engineering and I&CPS Europe, Genoa, Italy, June 2019.
[9] L. Nagel, R. Rohrer, "*Computer Analysis of Nonlinear Circuits, Excluding Radiation (CANCER)*", IEEE Journal on Solid-State Circuits, Vol. Sc-6, No.4, August 1971.
[10] L. Pileggi, R. Rohrer, C. Visweswariah, *Electronic Circuit & System Simulation Methods*, McGraw-Hill, Inc., New York, NY, USA, 1995.
[11] M. Jereminov, "Equivalent Circuit Programming," Doctoral thesis, Department of Electrical and. Computer Engineering., Carnegie Mellon University., Pittsburgh, PA, USA, August 2019.
[12] T. Chen, et.al., "On the Network Topology Dependent Solution Count of the Algebraic Load Flow Equations", IEEE Trans. on Power Syst, 2017.
[13] I. A. Hiskens, R. J. Davy, "Exploring the Power Flow Solution Space Boundary", IEEE Trans. on Power Systems, Vol. 16-3 August 2001.
[14] C. Liu, et.al. "Toward a CPFLOW-based algorithm to compute all the type-1 load-flow solutions in electric power systems," in *IEEE Trans. on Circuits and Systems I*: vol. 52, no. 3, pp. 625-630, March 2005.
[15] D. K. Molzahn, et.al., "Sufficient Conditions for Power Flow Insolvability Considering Reactive Power Limited Generators with Applications to Voltage Stability Margins", IREP, August 2013, Greece.
[16] T. J. Overbye, "A power flow measure for unsolvable cases", IEEE Transactions on Power Systems, vol. 9, no. 3, pp. 1359-1356, 1994.
[17] S. Yu, et.al., "Simple certificate of solvability of power equations for distribution systems", IEEE PES GM Denver, CO, USA, 2015.
[18] H. D. Nguyen, et.al., "Appearance of multiple stable load flow solutions under power flow reversal conditions", IEEE PES GM, 2014.
[19] A. Trias, "The holomorphic embedding load flow method", IEEE PES General Meeting, 2012, July 2012, pp. 1-8.
[20] I. Wallance, et.al. "Alternative PV Bus Modeling with Holomorphic Embedding Load Flow Method", ArXiv e-prints, July 2016.
[21] S.W. Director, R. Rohrer, "The Generalized Adjoint Network and Network Sensitivities", IEEE Trans. on Circuit Theory, vol. 16-3, 1969.
[22] S.W. Director, R, Rohrer, "Automated Network Design-The Frequency Domain Case", IEEE Trans. on Circuit Theory, vol. 16, no3, August 1969.
[23] M. Jereminov, A. Pandey and L. Pileggi, "Equivalent circuit formulation for solving AC optimal power flow," *IEEE Trans. on Power Systems*. DOI:10.1109/TPWRS.2018.2888907.
[24] M. Jereminov, et. al., "Equivalent Circuit Programming for Estimating the State of a Power System," PowerTech Milano, June 2019.
[25] D. Bromberg, M. Jereminov, L. Xin, G. Hug, L. Pileggi, "An Equivalent Circuit Formulation of the Power Flow Problem with Current and Voltage State Variables", *PowerTech Eindhoven, June 2015*.
[26] A. Pandey, et.al., "Unified Power System Analyses and Models using Equivalent Circuit Formulation," *IEEE PES ISGT*, USA, 2016.
[27] B. Stott, "Review of load-flow calculation methods," in Proceedings of the IEEE, vol. 62, no. 7, pp. 916-929, July 1974.
[28] A. Pandey, "Robust Steady-State Analysis of Power Grid using Equivalent Circuit Formulation with Circuit Simulation Methods," Doctoral Thesis, Carnegie Mellon University, August 2018.
[29] S. Boyd, L. Vandenberghe, *Convex Optimization*, Cambridge University Press, New York, NY, USA, 2004.
[30] M. Jereminov, A. Pandey, D. M. Bromberg, X. Li, G. Hug, L. Pileggi, "Steady-State Analysis of Power System Harmonics Using Equivalent Split-Circuit Models", ISGT Europe, Ljubljana, October 2016.
[31] M. Jereminov, D. M. Bromberg, A. Pandey, X. Li, G. Hug, L. Pileggi, "An Equivalent Circuit Formulation for Three-Phase Power Flow Analysis of Distribution Systems", *IEEE T&D Dallas, 2016.*
[32] Y. Wang, et.al., "Analysis of ill-conditioned power flow problems using voltage stability methodology", IEEET&D, Vol. 148-5, September 2001.
[33] R. Zimmerman, C. Murillo-Sanchez, R. Thomas, "MATPOWER: Steady-state operations, planning and analysis tools for power systems research and education", IEEE Trans. on Power Systems, vol. 26, no.1, Feb 2011.
[34] C. Josz, et.al, "AC Power Flow Data in MATPOWER and QCQP Format: iTesla RTE Snapshots and PEGASE". ArXiv e-prints, March 2016.
[35] A. B. Birchfield, et.al., "Power Flow Convergence and Reactive Power Planning in Creation of Large Synthetic Grids", IEEE Trans. on Power Systems, 2018.
[36] Transmission System Planning Performance Requirements, NREC Standard TPL -001-4, 2015.
[37] K.S. Kundert, A. Sangiovanni – Vincentelli, "Simulation of Nonlinear Circuits in the Frequency Domain", *IEEE Transactions on CAD*, Vol. CAD – 5, No.4, pp.521 – 535, October 1986.
[38] A. Semlyen, F. de Leon, "Quasi-Newton Power Flow using partial Jacobian updates," in *IEEE Transactions on Power Systems*, Vol. 16, No. 3, August 2001.
[39] D. Bromberg, A. Pandey, H. Zheng, Y. Li, "Robust Solution of High Renewable Penetration Planning Cases in SUGAR," Technical Conference on Increasing Real-Time and Day-Ahead Market Efficiency and Enhancing Resilience through Improved Software, FERC, 2019.